\documentclass[conference]{IEEEtran}

\usepackage[dvipsnames]{xcolor}
\usepackage{epsfig}
\usepackage{times}
\usepackage{float}
\usepackage{afterpage}
\usepackage{amsmath}
\usepackage{amstext,cite}
\usepackage{amssymb,bm}
\usepackage{latexsym} 
\usepackage{color}
\usepackage{graphicx}
\usepackage{amsthm}
\usepackage[center]{caption}
\usepackage{pstricks}
\usepackage{caption}
\usepackage{subcaption} 
\usepackage{booktabs}
\usepackage{multicol}
\usepackage{lipsum}%
\usepackage[T1]{fontenc}
\usepackage{hyperref}
\usepackage{aecompl}
\usepackage{mathrsfs}
\usepackage{bbm}
\usepackage{multirow}

\usepackage{tikz}
\usetikzlibrary{shapes.multipart}
\usetikzlibrary{automata}
\usetikzlibrary{fit}
\usetikzlibrary{shapes.geometric,positioning}
\usetikzlibrary{calc}
\usetikzlibrary{patterns.meta}
\usepackage{pgfplots}
\usetikzlibrary{plotmarks}
\usetikzlibrary{spy}

\usepackage{soul}
\usepackage{cancel}

\allowdisplaybreaks

\long\def\comment#1{}

\newcommand{\ZZ}{{\mathbb Z}}
\newcommand{\FF}{{\mathbb F}}

\newcommand{\EE}{{\mathbb E}}
\newcommand{\NN}{{\mathbb N}}

\newcommand{\dv}{{\mathbf d}}
\newcommand{\ev}{{\mathbf e}}

\newcommand{\Bm}{{\mathbf B}}
\newcommand{\Cm}{{\mathbf C}}
\newcommand{\Dm}{{\mathbf D}}
\newcommand{\Em}{{\mathbf E}}
\newcommand{\Fm}{{\mathbf F}}
\newcommand{\Gm}{{\mathbf G}}

\newcommand{\Id}{{\mathbf I}}

\newcommand{\Mm}{{\mathbf M}}

\newcommand{\Um}{{\mathbf U}}

\newcommand{\Vm}{{\mathbf V}}

\newcommand{\Ac}{{\mathcal A}}

\newcommand{\Gc}{{\mathcal G}}

\newcommand{\Ic}{{\mathcal I}}

\newcommand{\Lc}{{\mathcal L}}
\newcommand{\Mc}{{\mathcal M}}

\newcommand{\Pc}{{\mathcal P}}
\newcommand{\Qc}{{\mathcal Q}}
\newcommand{\Rc}{{\mathcal R}}
\newcommand{\Sc}{{\mathcal S}}
\newcommand{\Tc}{{\mathcal T}}

\newcommand{\Wc}{{\mathcal W}}

\newcommand{\msf}{{\mathsf m}}

\newcommand{\qsf}{{\mathsf q}}

\newcommand{\usf}{{\mathsf u}}

\newcommand{\Bsf}{{\mathsf B}}

\newcommand{\Msf}{{\mathsf M}}
\newcommand{\Nsf}{{\mathsf N}}

\newcommand{\Rsf}{{\mathsf R}}

\theoremstyle{definition}
\newtheorem*{rem*}{Remark}

\theoremstyle{plain}
\newtheorem{thm}{Theorem}%
\newtheorem{cor}{Corollary}
\newtheorem{lem}{Lemma}

\newtheorem{rem}{Remark}
\newtheorem{defn}{\protect\definitionname}

\providecommand{\definitionname}{Definition}

\newcommand{\Ksf}{{\mathsf K}}
\newcommand{\rk}{{\mathsf{rk}}}%
\newcommand{\aas}{\stackrel{\text{a.a.s.}}{=}}

\title{On Coded Caching Systems with Decentralized Linear Coding Placement}

\begin{document}

\author{
\IEEEauthorblockN{Yinbin Ma and Daniela Tuninetti \\}
\IEEEauthorblockA{University of Illinois Chicago, Chicago, IL 60607, USA \\ Email:\{yma52, danielat\}@uic.edu}
\thanks{The authors are with the Electrical and Computer Engineering Department of the University of Illinois Chicago, Chicago, IL 60607, USA (e-mail: yma52@uic.edu, danielat@uic.edu). 
Parts of this paper were presented in the 2022~\cite{ma2022coded}, in the 2023~\cite{ma2023demand}, and in the 2024~\cite{ma2024demand} IEEE International Symposium on Information Theory (ISIT). 
This work was supported in part by NSF Awards 1910309 and 2312229.}
}
\maketitle

\IEEEpeerreviewmaketitle

\begin{abstract}
Coded caching is a technique that leverages locally cached contents at the end users to reduce the network's peak-time communication load.  
Coded caching has been shown to achieve significant performance gains with a centralized placement orchestrated by the server and is thus considered a promising technique to boost performance in future networks by effectively trading off bandwidth for storage.
To tackle issues caused by the synchronized placement, previous works focused on decentralized placement and found the exact worst-case load with uncoded placement.
In this paper, we focus on a decentralized coded caching system with random linear coding placement,
and investigate the fundamental limits of a linear coding placement where each user independently and uniformly caches random linear coding symbols of a single file. %
We propose achievable and converse bounds on the worst-case load, which are shown to meet under certain conditions.
\end{abstract}


\section{Introduction}
\label{sec:intro}

Coded caching, first introduced by Maddah-Ali and Niesen in~\cite{maddah2014fundamental}, leverages locally cached contents at the users to reduce the communication load during peak-traffic times. A coded caching system has two phases. During the cache placement phase, the server populates the users' local caches, without knowing the users' future demands. During the delivery phase, the server broadcasts coded multicast messages to satisfy the users' demands. It is a promising technique as coded delivery reduces communication load significantly, with a well-designed placement strategy, compared to uncoded delivery.

The achievable scheme proposed in~\cite{maddah2014fundamental} (referred to as MAN in the following) has a combinatorial {\it uncoded cache placement} phase\footnote{Uncoded cache placement means that bits of the files are directly copied into the caches without any coding.} and a network coded delivery phase. In~\cite{yu2017exact}, an improved delivery was proposed (referred to as YMA in the following), which improves on the MAN delivery by removing those linearly dependent multicast messages that may occur when a file is requested by multiple users. The scheme with MAN placement with YMA delivery matches the converse bound derived in~\cite{wan2020index} under the constraint of uncoded placement~\cite{yu2017exact}; and it is otherwise order optimal to within a multiplicative factor of two if no restrictions are imposed on the placement phase~\cite{yu2018characterizing}.

Wan {\em et al.} in~\cite{wan2021optimal} extended the single-file retrieval model in~\cite{maddah2014fundamental} so as to include Scalar Linear Function Retrieval (SLFR), where users can demand a linear combination of the files. 
By generalizing the YMA scheme,~\cite{wan2021optimal} proved that SLFR attains the same load as for single-file retrieval under uncoded placement. A general construction for SLFR with uncoded placement and linear coding for the delivery can be found in~\cite{ma2021general}.

For the original coded caching model~\cite{maddah2014fundamental}, coded placement is known not only to strictly improve performance compared to uncoded placement, but also to be exactly optimal in some regimes:
\cite{chen2016fundamental} shows how to achieve the cut-set bound in the small memory regime when there are more users than files; 
\cite{gomez2018fundamental} generalizes the scheme in~\cite{chen2016fundamental} and extends the memory range where the optimal performance is known; 
\cite{tian2018caching} proposes a scheme based on interference elimination for the case of more users than files;
\cite{gomez2018fundamental} shows an improved performance compared to~\cite{tian2018caching} in the small memory regime when there are more users than files;
\cite{tian2018symmetry} derives the optimal performance for the case of two users (and any number of files), and a partial characterization for the case of two files (and any number of users).

For converse bounds without any restrictions on the placement phase and that improve on the cut-set bound in~\cite{maddah2014fundamental}, in addition to those in~\cite{yu2018characterizing}, %
\cite{wang2018improved} proposes a converse bound where the worst-case and average-case loads are within a multiplicative gap of 2.315 and 2.507 to the YMA scheme respectively;
we note that~\cite{sengupta2017improved} derived a converse bound tighter than the cut-set bound, and~\cite{tian2018symmetry} proposed a computer-aided methodology to derive converse bounds based on Shannon-type inequalities; the method of~\cite{tian2018symmetry} is unfortunately limited to small instances of the problem, as the computational complexity of the resulting linear program scales doubly exponentially with the number of users.

The aforementioned achievable schemes, including the YMA scheme, are based on the original MAN system. The limitation of the MAN system is that the placement phase must be a synchronized process orchestrated by a centralized server when all users are present.
That is, the local caches of all users are jointly distributed random variables, and each cache may depend on the others.
In a practical system, such as a wireless system, users may join or leave the network at any time, so such synchronization may not be possible.
A novel approach to tackle this issue is the decentralized placement phase. That is, rather than a joint distribution, we assume each local cache is an independent and identically distributed (i.i.d.) random variable depending only on the library.
\cite{maddah2014decentralized} introduces an uncoded cache placement phase, where each user caches each symbol of the library independently, uniformly, and randomly.
The achievability consists of multiple rounds of MAN delivery and attains order optimality.
A following work~\cite{yu2017exact} improves the achievability with YMA delivery and proves it is optimal under the constraint of uncoded placement; it was later proved to be order-optimal within a multiplicative factor of~2 for any general placement~\cite{yu2018characterizing}. 
Rather than uncoded placement,
\cite{wei2017novel,reisizadeh2018erasure} use Maximum Distance Separable (MDS) codes on each file and then proceed with the decentralized scheme~\cite{yu2017exact} over these coded files.
\cite{reisizadeh2018erasure} showed that it achieves the same load performance as the centralized YMA scheme under certain conditions.

\subsection{Contributions}
The goal of this paper is to shed light on {\it the decentralized coded caching system with linear coding placement}.
We formally propose a decentralized coded caching system with a linear coding prefetching $P_\Em$, which is a probability distribution used to generate the random linear coding symbols of the library.
We define the placement as consisting of, independently and uniformly, random linear coding symbols of at most $\msf$ files, referred to as $\msf$-LinP.
We characterize the worst-case load of $1$-LinP and formulate a linear program in~\eqref{eq:lpachievability} to derive the achievability.
We show that exact optimality under the constraint of $1$-LinP placement, summarized in Table~\ref{tab:generalsolutions}, is achievable for at most $3$ users, and otherwise in some memory regimes for at least $4$ users.

\subsection{Paper Outline}
The rest of the paper is organized as follows.
Section~\ref{sec: problem formation} formulates the decentralized caching problem.
Section~\ref{sec:achievability} summarizes the achievability for $1$-LinP and examples.
Section~\ref{sec:proofofHT1} and~\ref{sec:proofofconverse} are proofs of our results.
Section~\ref{sec:conclusion} concludes the paper.

\subsection{Notation Convention}
\label{sec: notation}
We adopt the following notation convention.
\begin{itemize}
    \item Calligraphic symbols denote sets, bold lowercase symbols denote 
    vectors, bold uppercase symbols denote matrices, and sans-serif symbols denote 
    system parameters.
    \item $\Mm[\Qc]$ denotes the submatrix of $\Mm$ obtained by selecting the rows indexed by $\Qc$.
    Similarly, $\dv[\Ic]$ is the subvector of $\dv$ obtained by selecting the elements indexed by $\Ic$.
    \item For integers $a$ and $b$, we let $[a: b] := \{a, a+1, \ldots, b\}$, and $[a] := \{1, 2, \ldots, a\}$.
    \item For sets $\Sc$ and $\Qc$, we let $\Sc \setminus \Qc := \{k: k \in \Sc, k \notin \Qc\}$.
    \item For a vector $\dv$, $\mathsf{rank}(\dv)$ is the number of distinct elements in $\dv$. %
    \item For a matrix $\Dm$, $\mathsf{rank}(\Dm)$ is the rank of $\Dm$. %
    \item For a collection $\{Z_1, \ldots Z_n\}$ and an index set $\Sc \subseteq [n]$, we let $Z_\Sc := \{Z_i: i \in \Sc\}$.
    \item For a ground set $\Gc$ and an integer $t$, we let $\Omega_{\Gc}^{t} := \{ \Tc \subseteq \Gc : |\Tc| = t\}$. 
    \item For a set $\Sc$, the power set of $\Sc$ is $\Pc(\Sc)$.
    \item For integers $a$ and $b$, $\binom{a}{b}$ is the binomial coefficient, or zero if  $a \geq b \geq 0$ does not hold.
    \item The $i$-th standard basis vector is the vector $\ev_{i}$ that has only one non-zero component equal to 1 in position $i$.
    \item For a real number $x$, we let $[x]^{+} := \max(0, x)$.
    \item The Kronecker product is denoted by $\otimes$.
    \item For a positive integer $m$, we denote $\Id_m$ as the identity matrix with the dimension $m \times m$.
    \item Given a random variable $X$, we define the entropy of $X$ as $H(X)$; the choice of base for the log depends on the context.  
  \end{itemize}

\section{Problem Formulation}
\label{sec: problem formation}

A decentralized coded caching system, given a prefetching scheme $P_\Em$, consists of the following.
\begin{itemize}
    \item A central server stores $\Nsf$ files, denoted as $F_1, \ldots, F_\Nsf$. 
    \item Each file has $\Bsf$ i.i.d. uniformly distributed symbols over a finite field $\mathbb{F}_\qsf$, where $\qsf$ is a prime-power number.
    \item The server communicates with users through an error-free shared link.
    \item Each user $k \in \ZZ^+$ has a local cache that can contain up to $\Msf \Bsf$ symbols, denoted by $Z_k$, where $\Msf \in [0, \Nsf]$.
    We refer to $\Msf$ as the {\it memory size}, and denote $\gamma := \Msf/\Nsf \in [0, 1]$ as the {\it memory ratio}.
    \item The prefetching scheme $P_\Em$ is a probability distribution over $\FF_\qsf^{\Nsf \Bsf \times \Msf \Bsf}$. More specifically, $P_\Em$ is a probability measure on $\FF_\qsf^{\Nsf\Bsf \times \Msf\Bsf}$.
    \item The server sends the signal $X$ to $\Ksf$ active users through the shared link, where $\Ksf \in \ZZ^+$ and $X$ has no more than $\Rsf \Bsf$ symbols, with $\Rsf\in [0, \Nsf]$.
    We refer to $\Rsf$ as the {\it load}.

    \item The server has a {\it placement phase} and a {\it delivery phase}.

    The placement phase occurs when a new user connects to the server for the first time. The server initializes the cache content of the new user. During the placement phase, the server does not have any prior knowledge of the total number of users, the identity of active users, or which files will be requested in the future.
    Each user then requests a scalar linear function of all files from the server at any time after the placement phase is done, then waits to receive messages from the server. 
    The server has a queue to store the demands from the active users. The delivery phase occurs when the server decides to serve all active users. 

    \item 
    Placement Phase: Let $F := [F_1; \ldots; F_\Nsf] \in \mathbb{F}_\qsf^{\Nsf\Bsf}$ be the column vector that concatenates all files. For a new user indexed by $k \in \NN$, the server generates an i.i.d. random matrix $\Em_k \in \FF_\qsf^{\Nsf \Bsf \times \Msf \Bsf}$ following the probability distribution $P_{\Em}$, then populates the caches as a function of all files, 
    \begin{align}
        Z_k = \Em_k^T F \in \FF_\qsf^{\Msf \Bsf},
        \label{eq:cacheconstraint}
    \end{align}

    \item 
    Delivery Phase: Without loss of generality, we assume the first $\Ksf$ users are active, that is,
user $k \in [\Ksf]$ generates a vector $\dv_k \in \FF_\qsf^{\Nsf}$, then it 
demands a scalar linear function of files,
\begin{align}
	B_k := \sum_{n\in [\Nsf]} d_{k,n} F_n, \quad \forall k\in[\Ksf],
    \label{eq:SLFRdemand}
\end{align}
where $\dv_k = [d_{k,1},\ldots,d_{k,\Nsf}]$.
Let the demand matrix $\Dm = [\dv_1; \ldots; \dv_\Ksf] \in \FF_\qsf^{\Ksf \times \Nsf}$,
the server sends the message $X$ as a function of the files and of the demands of the active users, i.e., 
\begin{align}
    &H(X \mid \Dm, F) = 0. %
    \label{eq:encoding}
\end{align}

    \item 
    Decoding: For every $\epsilon > 0$, each active user $k \in [\Ksf]$ must be able to decode its desired linear function $B_k$ with a probability of error at most $\epsilon$, i.e.,
\begin{align}
&H(\hat{B}_{k} \mid Z_k, X, \Dm) = 0, \ \mathsf{Pr}(\hat{B}_{k} \neq B_{k}) \leq \epsilon.
\label{eq:decoding} 
\end{align}
    \item Performance:
For $\gamma \in [0, 1]$, equivalently when $\Msf \in [0, \Nsf]$, %
we define $\Rsf^\star$ as the minimal {\it worst-case} load given $(\Em_1, \ldots, \Em_\Ksf)$, that is,
\begin{align*}
    \Rsf^\star &= \min_{X} \max_{\Dm} \{\Rsf: 
\textrm{\small conditions}
\notag \\ &\quad
\textrm{in~\eqref{eq:cacheconstraint}, \eqref{eq:SLFRdemand}, \eqref{eq:encoding} and~\eqref{eq:decoding} are satisfied}\}.
\end{align*}
the expected worst-case load under the prefetching scheme $P_\Em$ when serving $\Ksf$ users is defined as
\begin{align}
&\Delta^{(P_\Em)}_\Ksf(\gamma) = \limsup_{\Bsf \rightarrow \infty} \ \EE_{P_\Em } \, \Rsf^\star(\Em_1, \ldots, \Em_\Ksf).
\label{eq:load}
\end{align}
When the context is clear, we omit the subscript $\Ksf$ and simplify~\eqref{eq:load} as $\Delta^{(P_\Em)}$.
\end{itemize}

We stress that the server and users don't have any prior assumptions on $\Ksf$, the number of active user in the delivery phase. For~\eqref{eq:load}, we discuss the infinite dimensional vector where each $j$-th element represents $\Delta^{(P_\Em)}_j$ in later Remark~\ref{rem:uncodedrateregion}.

\begin{defn} \label{def:linp}
    Given a user $k \in \ZZ^+$, fix $\msf \in [\Nsf]$, %
    let $\Mc \in \Omega_{[\Nsf]}^{\msf}$ be a subset of $[\Nsf]$ with size $\msf$. The server then populates the cache content for user $k$ as
    \begin{align}
        Z_k = \left\{ \sum_{n \in \Mc} \Em_{k,\Mc,n} F_n: \Mc \in \Omega_{[\Nsf]}^{\msf} \right\}, \label{eq:QLinPcache}
    \end{align}
    where $\Em_{k,\Mc,n} \in \FF_\qsf^{\Bsf \times \Msf_\msf \Bsf}$ is chosen uniformly at random. 
    Equivalently, $\Em_k$ is a linear projection of at most $\msf$ files.  
    We define the support of such a prefetching scheme as $\msf$-LinP, denoted by $\mathfrak{L}_\msf$, that is, $P(\Em_k \in \mathfrak{L}_\msf) \aas 1$.
\end{defn}
Definition~\ref{def:linp} implies that,
\begin{align}
    \mathfrak{L}_1 \subseteq \mathfrak{L}_2 \subseteq \ldots \subseteq \mathfrak{L}_\Nsf. \label{eq:mlinpineq}
\end{align}
Since every $\Em_{k,\Mc,n}$ in~\eqref{eq:QLinPcache} is almost linearly independent, then
$$\rk(\Em_k)/\Bsf = \binom{\Nsf}{\msf} \Msf_\msf \aas \Msf .$$
We stress \emph{the main reason for investigating the tradeoff between $\mathfrak{L}_\msf$ and $\Delta_{\Ksf}^{(\mathfrak{L}_\msf)}$ for every $\msf \in [\Nsf]$ separately.}
Suppose $\Em_k$ is drawn  from $\mathfrak{L}_\msf$ where $\msf \geq 2$, despite the fact of~\eqref{eq:mlinpineq} implies the nested structure, we still have $P(\Em_k \in \mathfrak{L}_\usf) \aas 0$ for any $1 \leq \usf < \msf$. In other words, $\Delta_{\Ksf}^{(\mathfrak{L}_\msf)}$ concentrates on the load where $\Em_k \in \mathfrak{L}_\msf$, regardless other smaller $\mathfrak{L}_\usf$ spaces.

\begin{rem}
    \label{rem:memorysharingimpossible}
    Memory sharing is impossible for all $m$-LinP. More specifically, we first partition each file into two disjoint parts, $F_n = [F_{n, 1}, F_{n, 2}]$, where $F_{n,i} \in \FF_\qsf^{\alpha_i \Bsf}$ and $\alpha_1 + \alpha_2 = 1$. Given any two satisfying $\msf$-LinP placements for user~$k \in \ZZ^+$, %
    \begin{align*}
        Z^{(i)}_k = \left\{ \sum_{n \in \Mc} \Em^{(i)}_{k,\Mc,n} F_{n,i}: \Mc \in \Omega_{[\Nsf]}^{\msf} \right\} \in \FF_\qsf^{\alpha_i\Msf\Bsf}, \ i \in \{1,2\}
    \end{align*}
    where $\Em^{(i)}_{k,\Mc,n} \in \FF_\qsf^{\alpha_i \Bsf \times \alpha_i \Msf_\msf \Bsf}$. %
    The $\Em_{k,\Mc,n}$ of the joint placement $Z_k := Z^{(1)}_k \cup Z^{(2)}_k$ defined in~\eqref{eq:QLinPcache}, is %
    \begin{align*}
        \Em_{k,\Mc,n} := \begin{bmatrix}
            \Em^{(1)}_{k,\Mc,n} & 0 \\ 0 & \Em^{(2)}_{k,\Mc,n}
        \end{bmatrix}, %
    \end{align*}
    that is, the off-diagonal sub-blocks of $\Em_{k,\Mc,n}$ are not randomly chosen when $\alpha_1\alpha_2 > 0$. Thus, $Z_k$ attained by memory sharing, is not qualified as an $\msf$-LinP. %

\end{rem}

Similar to the classical coded caching system~\cite{tian2018symmetry}, for the decentralized coded caching with any prefetching $P_\Em$, $\Delta^{(P_\Em)}$ is unchanged if we permute the index of users or files, that is, both {\it user symmetry} and {\it file symmetry} are preserved.

\subsection{Centralized Scheme with Uncoded Placement}
\label{sec:YMA}
For the classical coded caching model, the YMA-type scheme is optimal under the constraint of uncoded placement, both for single-file~\cite{yu2017exact} and for scalar linear function retrieval~\cite{wan2021optimal} demands.
We describe next the YMA scheme with SLFR demands, as it will be needed to introduce the decentralized scheme in later sections.

\paragraph*{Placement Phase} 
Fix $t \in [0: \Ksf]$.
Partition each file into $\binom{\Ksf}{t}$ equal-size subfiles as
\begin{align}
    F_i = (F_{i,\Wc}  : \Wc \in \Omega_{[\Ksf]}^{t}),  \quad \forall i \in [\Nsf].
    \label{eq:MANsplit}
\end{align}
The cache content of user~$k$ is
\begin{align}
    Z_k = (F_{i,\Wc} : i \in [\Nsf], \Wc \in \Omega_{[\Ksf]}^{t}, k \in \Wc), \quad \forall k \in [\Ksf].
    \label{eq:MANacahe}
\end{align}
The needed memory size is
\begin{align}
	\Msf^\text{\rm(YMA)}_t := \Nsf \frac{\binom{\Ksf-1}{t-1}}{\binom{\Ksf}{t}} = \Nsf \frac{t}{\Ksf}.
	\label{eq:MANmemorysize}
\end{align}
The placement in~\eqref{eq:MANacahe} is referred to as {\it centralized placement} because it requires coordination among users during the placement phase~\cite{maddah2014fundamental}.

\paragraph*{Delivery Phase} 
Given the file partitioning in~\eqref{eq:MANacahe}, we can partition~\eqref{eq:SLFRdemand} as $B_k = (B_{k,\Wc}  : \Wc \in \Omega_{[\Ksf]}^{t})$ where the {\it block} $B_{k,\Wc}$ is defined as
\begin{align}
	B_{k,\Wc} := \sum_{n\in [\Nsf]} d_{k,n} F_{n,\Wc}, \quad \forall k \in [\Ksf].
    \label{eq:SLFRblock}
\end{align}
Given the demand matrix $\Dm$ defined in~\eqref{eq:SLFRdemand}, the server constructs the multicast messages 
\begin{align}
    X_\Sc = \sum_{k \in \Sc} \alpha_{k, \Sc \setminus \{k\}} \, B_{k, \Sc \setminus \{k\}}, \ \forall \Sc \in \Omega_{[\Ksf]}^{t+1},
    \label{eq:MANmm}
\end{align}
where the {\it encoding coefficients} $\alpha_{k, \Sc \setminus \{k\}} \in \{+1,-1\}$  are selected as shown in~\cite{wan2021optimal} (for a more general construction, we refer the reader to~\cite{ma2021general}) %
, for all $\Sc \in \Omega_{[\Ksf]}^{t+1}$ and $k \in [\Ksf]$.
The server selects a {\it leader set} $\Lc \subseteq [\Ksf]$ such that $\{d_k : k \in \Lc\}$ are all distinct, and sends
\begin{align}
    X = (\Lc, \Dm, X_\Sc: \Sc \in \Omega_{[\Ksf]}^{t+1}, \Sc \cap \Lc \neq \emptyset ).
    \label{eq:YMAmm}
\end{align}

\paragraph*{Decoding}
As shown in~\cite{wan2021optimal,ma2021general}, the delivery signal in~\eqref{eq:YMAmm} allows all users to successfully decode their requested SLFR demand. The key observation is that it is possible to choose the encoding coefficients in~\eqref{eq:MANmm} in such a way that the non-leader users, indexed by $[\Ksf] \setminus \Lc$, can locally reconstruct the not-sent multicast messages $X_\Ac,$ for all $\Ac \in \Omega_{[\Ksf] \setminus \Lc}^{t+1},$ from the delivery signal in~\eqref{eq:YMAmm}. 
Note that user $k \in \Sc$ can recover the missing block $B_{k, \Sc \setminus \{k\}}$ from $X_\Sc$ in~\eqref{eq:MANmm} by ``caching out'' $\sum_{u \in \Sc \setminus \{k\}} B_{u, \Sc \setminus \{u\}}$, which can be computed from $Z_k$ in~\eqref{eq:MANacahe}.

\paragraph*{Performance}
For the delivery signal in~\eqref{eq:YMAmm} with $|\Lc|$ leaders we have an upper bound on $H(X)$
\begin{align}
|\Lc|\log_\qsf(\Ksf)+\Ksf\Nsf + \frac{ \binom{\Ksf}{t+1} - \binom{\Ksf-|\Lc|}{t+1}}{ \binom{\Ksf}{t} }\Bsf. %
\end{align}
Thus, for a large $\Bsf$ and $|\Lc| = \mathsf{rank}(\Dm)$, %
the lower convex envelope of the following points is achievable
\begin{align}
    ( \gamma_t, \Rsf_t )^\text{\rm{(YMA)}}  = \biggl(
    \frac{t}{\Ksf}, %
    \frac{\binom{\Ksf}{t+1} - \binom{\Ksf -|\Lc|}{t+1}}{\binom{\Ksf}{t}} \biggr), %
    \label{eq:performanceYMA}
\end{align}
where $t \in [0: \Ksf]$.
The YMA scheme is optimal under the constraint of uncoded placement for the centralized coded caching system. %

\subsection{Decentralized Scheme with Uncoded Placement}
\label{sec:DecentralizedPlacement+hotplug}
The decentralized scheme refers to the case where users cache each symbol of the library in an i.i.d. fashion with probability $\gamma$~\cite{maddah2014decentralized,yu2017exact}, %
that is, for every $k \in \NN^+$, every column vector in $\Em_k$ is a unit vector, and chosen from $\Id_{\Nsf\Bsf}$ uniformly at random.

When the server knows the demands $\Dm$ from $\Ksf$ active users and decides to start the delivery phase, for every $n \in [\Nsf]$ and $s \in [0: \Ksf]$, fix $\Sc \in \Omega_{[\Ksf]}^{s}$, we denote $F_{n,\Sc}$ as the set of bits cached by $\Sc$ users only.
When $\Bsf$ goes to infinity, $|F_{n,\Sc}|/\Bsf$ converges to a constant, denoted by $\beta_s$, in probability regardless of $n$ and $\Sc$, that is,
\begin{align}
    \beta_s := |F_{n,\Sc}|/\Bsf \aas \gamma^{s} (1-\gamma)^{\Ksf-s}. \label{eq:defbetas}
\end{align}
Note that when $s=0$, it results in $\Sc = \emptyset$, and $F_{n, \emptyset}$ means those symbols of $F_n$ which are never cached by any of the $\Ksf$ users; when $s=\Ksf$, it results in $\Sc = [\Ksf]$, and $F_{n, [\Ksf]}$ means those symbols of $F_n$ which are cached by all users.
The server first sends $\{F_{n, \emptyset}: n \in [\Nsf]\}$. Then for every $s \in [\Ksf-1]$, the server sends the YMA-type multicast messages and exchanges these subfiles $\{F_{n, \Sc}: n \in [\Nsf], \Sc \in \Omega_{[\Ksf]}^{s}\}$ among users. The total communication load is thus,
\begin{align}
    \Rsf^{\rm (U)} &\aas \sum_{s=0}^{\Ksf-1} \ \beta_s \ \left(\binom{\Ksf}{s+1} - \binom{\Ksf-|\Lc|}{s+1}\right) \\ 
    &= \frac{1-\gamma}{\gamma}\Big( 1-(1-\gamma)^{|\Lc|} \Big),
\end{align}
where $|\Lc| = \mathsf{rank}(\Dm) \leq \min(\Ksf, \Nsf)$.

\begin{thm}[Exact load on uncoded placement~\cite{yu2017exact}]
    \label{thm:extensionYMAdecentralized}
    For a decentralized coded caching system with $\Ksf$ users and $\Nsf$ files under the constraint of uncoded placement, the exact expected worst-case load is
    \begin{align}
        \Delta^{\rm (U)}_\Ksf(\gamma) 
    = \frac{1-\gamma}{\gamma}\Big( 1-(1-\gamma)^{\min(\Ksf,\Nsf)} \Big),
    \label{eq:performanceDecentralized} 
    \end{align}
    where $\gamma \in (0, 1]$. When $\gamma = 0$, $\Delta^{\rm (U)} = \min(\Nsf, \Ksf)$.
\end{thm}

\begin{rem} \label{rem:uncodedrateregion}
Theorem~\ref{thm:extensionYMAdecentralized} is equivalent to the infinite dimensional region $\Rc_{\text{peak}}$ in term of peak rate~\cite[Corollary~2]{yu2017exact}, which is defined as
\begin{align}
    \Rc_{\text{peak}} = \{x_k: k \in \ZZ^+, x_k \geq \Delta^{\rm (U)}_k\}.
\end{align}
Essentially $\Rc^{\text{peak}}$ depends on $\Nsf$ and $\gamma$.
\end{rem}

\subsection{Decentralized Scheme with Coded Placement}
Based on the decentralized scheme with uncoded placement, \cite{reisizadeh2018erasure, wei2017novel} consider an additional phase, named the precoding phase, by applying the Maximum Distance Separable (MDS) code on each file before the placement phase. That is,  each individual file is coded with a selected MDS rate $\theta \in (0, 1]$,
\begin{align}
    C_{n} := \Gm^T F_n, \ \text{where } \Gm \in \FF_\qsf^{\Bsf \times \Bsf /\theta}.
\end{align}
\begin{subequations}
We assume that $\qsf$ is large enough so such an MDS matrix $\Gm$ exists. Let $\gamma' := \Msf \theta/\Nsf$. Each user caches a subset of $\gamma' \Bsf$ symbols from every $C_n$; that is, each symbol is chosen independently by users uniformly at random with a probability $\gamma'$. The server then follows the same delivery phase as in the uncoded case.
However, not all multicast messages are delivered; otherwise, users would receive redundant symbols. Let $s' \in [\Ksf-1]$ and $\eta' \in (0, 1]$ be the unique solutions to
\begin{align}
    \eta' \binom{\Ksf-1}{s'} \beta'_{s'} + \sum_{s = s'+1}^{\Ksf} \binom{\Ksf-1}{s} \beta'_s = 1-\gamma, \label{eq:mdsdecendecoding}
\end{align}
where $\beta'_s := (\gamma')^{s} (1-\gamma')^{\Ksf-s}/ \theta$. Then the load is:
\begin{align}
    R^\text{\rm{(MDS)}} \aas \min_{\theta} \ \eta' c(s') \beta'_{s'} + \sum_{s = s'+1}^{\Ksf} c(s) \beta'_s,
\end{align}
where $c(t) := \binom{\Ksf}{t+1} - \binom{\Ksf-\min(\Ksf,\Nsf)}{t+1}$. That is, the server skips multicast messages benefiting fewer than $s'$ users and sends partial symbols of multicast messages benefiting exactly $s'$ users.
After choosing the fine-tuned $\theta^\star$, it is shown that when 
\begin{align*}
    0 \leq \Msf \leq \Nsf/\Ksf \text{ or } \frac{\Nsf}{1+\frac{(\Ksf-2)^{\Ksf-2}}{(\Ksf-1)^{\Ksf-1}}} \leq \Msf \leq \Nsf,
\end{align*}
$R^\text{\rm{(MDS)}}$ is exactly the load performance of the centralized scheme with uncoded placement, i.e., $R^\text{\rm{(MDS)}}=R^\text{\rm{(YMA)}}$.
\label{eq:mdsdecendlp}
\end{subequations}

\begin{rem} \label{rem:mds}
    We note that the fine-tuned $\theta^\star$ implicitly depends on $\Ksf$, which means that $\theta^\star$ may not be uniform for various $\Ksf$. 
    For example, if we fix $\gamma=0.25$ and $\Nsf=3$, then $\theta^\star \approx 0.067$ when $\Ksf=3$, and $\theta^\star \approx 0.588$ when $\Ksf=6$\footnote{This is an approximation only, as~\eqref{eq:mdsdecendlp} is not a convex optimization.}.
    However, the MDS rate, $\theta$, is determined in the placement phase and cannot be varied during the delivery phase.
    Therefore, $R^\text{\rm (MDS)}$ is not always feasible in our proposed model, as we do not restrict the range of the number of active users $\Ksf$.
\end{rem}

\section{Achievability with $1$-LinP placement}
\label{sec:achievability}
In this section, we provide an example of the decentralized system under the constraint of a linear coding placement of at most $1$ file (i.e., $1$-LinP) when $3$ users and $3$ files are present. 
Then we present the achievability of the decentralized scheme with $1$-LinP in Theorem~\ref{thm:HT1}. 

\subsection{Preliminary Results}
Before we present the proof of Theorem~\ref{thm:HT1}, we first present some preliminary results. %
\begin{lem}{{\cite[Lemma~1]{yao2024generic}}}
\label{lem:randomindependent}
Let $\Em \in \FF_\qsf^{n \times m}$. If the elements of $\Em$ are chosen i.i.d. uniformly from $\FF_\qsf$, then $\rk(\Em) \aas \min(m, n)$ as $\qsf \rightarrow \infty$.
\end{lem}
\begin{cor} \label{cor:randomindependent}
Let $\Em \in \FF_p^{n \Bsf \times m \Bsf}$. If the elements of $\Em$ are chosen i.i.d. uniformly from $\FF_p$, then $\rk(\Em)/\Bsf \aas \min(m, n)$ as $\Bsf \rightarrow \infty$ due to finite field extension.
\end{cor}

\begin{lem} \label{lem:sumoftwomatrix}
    Given any two matrices $\Mm_1 \in \FF_\qsf^{n \times m_1}$, and $\Mm_2 \in \FF_\qsf^{n \times m_2}$, we have
    \begin{multline*}
        \rk(\Mm_1) + \rk(\Mm_2) = \rk(\Mm_1 \cup \Mm_2) + \rk(\Mm_1 \cap \Mm_2).
    \end{multline*}
\end{lem}

\begin{figure}
    \center
    \scalebox{0.8}{
    \begin{tikzpicture}
    \node [draw,
        circle,
        minimum size =3cm, color = olive, thick] (U1) at (0.9,1.5){};
    \node [draw,
        circle,
        minimum size =3cm, color = magenta, thick] (U2) at (0,0){};
    \node [draw,
        circle,
        minimum size =3cm, color = teal, thick] (U3) at (1.8,0){};
    \node [draw, minimum width=6cm, minimum height=6cm, thick, rounded corners] (R) at (0.9, 0.4) {};

    \node at (0.9,0.4) {$\Em_{123}$};
    \node at (0,1) {\footnotesize $\Em_{12}$};
    \node at (1.8,1) {\footnotesize $\Em_{13}$};
    \node at (0.9,-0.5) {\footnotesize $\Em_{23}$};
    \node at (0.9,2.2) {$\Em_{1}$};
    \node at (-0.6,-0.4) {$\Em_{2}$};
    \node at (2.4,-0.4) {$\Em_{3}$};
    \node at (0.9, -2) {$\Id_\Bsf$};

    \end{tikzpicture}
    }
    \caption{\small The Venn diagram for $\langle \Em_1\rangle, \langle \Em_2\rangle, \langle \Em_3\rangle$. The joint space $\langle [\Em_1, \Em_2, \Em_3]\rangle$ is covered by the ground space $\Id_\Bsf$.}
    \label{fig:venn1}
\end{figure}
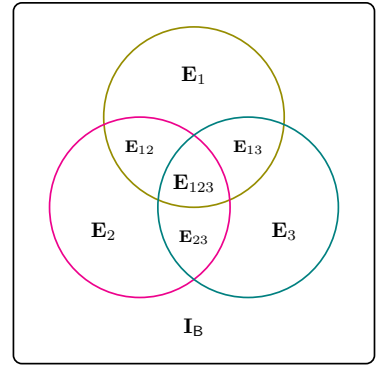

Given $\Ksf$ random matrices $\{\Em_1, \ldots, \Em_\Ksf\}$, 
every $\Em_k$ %
is independently chosen from $\FF_\qsf^{\Bsf \times \gamma \Bsf}$ uniformly at random, where $\gamma \in [0, 1]$.
for every $\Sc \subseteq [\Ksf]$
we define the intersection, union, and complement operations, respectively
\footnote{To keep the notation light, when the context is clear, we simplify the set as letters, for example, we write $\Sc = \{i,j\}$ as $ij$.}
,
\begin{align*}
    \Em_{(\Sc)} := \bigcup_{k \in \Sc} \langle \Em_k \rangle, \, %
    \Em_{\Sc} := \bigcap_{k \in \Sc} \langle \Em_k \rangle, \, 
    (\Em_{\Sc})^{C} := \langle \Id_\Bsf \rangle \setminus \langle \Em_\Sc \rangle.
\end{align*}
For convention, we define $\Em_{\emptyset} = \Id_\Bsf$.
By definition, $\Em_{\Sc}$ is the subspace spanned by the intersection of the column spaces of $\Em_k$ for all $k \in \Sc$. 
Fig.~\ref{fig:venn1} shows the Venn diagram for $\Ksf=3$ random matrices.  
By Corollary~\ref{cor:randomindependent}, when $\Bsf \rightarrow \infty$, for every $\Sc \subseteq [\Ksf]$, $s = |\Sc|$, we have $\rk(\Em_{(\Sc)})/\Bsf \aas \min(s\gamma, 1)$, and
\begin{align*}
    \rk(\Em_{\Sc})/\Bsf &= \rk\left((\bigcup_{k \in \Sc} (\Em_k)^C)^{C}\right)/\Bsf \aas \left[1 - s(1-\gamma)\right]^{+}. \label{eq:rankofintersection}
\end{align*}

\subsection{Example for $\Ksf = \Nsf = 3$}
\label{sec:K=N=3 example}

\paragraph*{Placement Phase}
By the definition of $1$-LinP in Definition~\ref{def:linp}, given $\gamma \in [0, 1]$,
for every user $k\in[3]$, the server generates the cache content of user~$k$ as follows:
\begin{align*}
    Z_k = \{ \Em_{k,n}^T F_n: n \in [3] \},
\end{align*}
where $\Em_{k,n} \in \FF_\qsf^{\Bsf \times \gamma\Bsf}$ is chosen i.i.d. uniformly at random. 

Given $n \in [3]$, $\Sc \subseteq [3]$ and $s := |\Sc|$, let $\Em_{\Sc,n} = \cap_{k \in \Sc} \langle \Em_{k,n}\rangle$.
By~\eqref{eq:rankofintersection}, the normalized rank of $\Em_{\Sc,n}$ should concentrate on $\tau_s$,
\begin{align*}
    \tau_s = \rk(\Em_{\Sc,n})/\Bsf \aas \left[1 - s(1-\gamma)\right]^+,
\end{align*}
where $\tau_0 = 1$.
Similar to the decentralized scheme in Subsection~\ref{sec:DecentralizedPlacement+hotplug}, 
We denote the coded subfiles computable by users in $\Sc \subseteq [3]$ \textbf{only} as $C_{n,\Sc} := \Vm_{\Sc, n}^T F_n, \text{where } \Vm_{\Sc, n} \in \FF_\qsf^{\Bsf \times \lambda_s \Bsf}$,
where the dimension of $\Vm_{\Sc,n}$ depends on $s$ due to file and user symmetry.
By definition, every $\Vm_{\Sc, n}$ is a subspace of $\Em_{\Tc, n}$ where $\Tc \subseteq \Sc$, for example,
\begin{align*}
    \Vm_{12, 1} \subseteq \Em_{12, 1} \subseteq \Em_{1, 1} \subseteq \Em_{\emptyset, 1}.
\end{align*}
Intuitively, $\Vm_{\Sc, n}$ is the basis that forms the subspace unique to the intersection $\Em_{\Sc,n}$.
Given any subset $\Tc \subseteq [3]$,
\begin{align*}
    \rk(\Em_{\Tc, n}) \geq \sum_{\Sc \subseteq [3], \Tc \subseteq \Sc} \rk(\Vm_{\Sc, n}).
\end{align*}
Equivalently, we have:
\begin{subequations}
\begin{align}
    \lambda_3 &\leq \tau_3 = \left[3\gamma - 2\right]^+, \\ 
    \lambda_2 + \lambda_3 &\leq \tau_2 =  \left[2\gamma - 1\right]^+, \\
    \lambda_1 + 2 \lambda_2 + \lambda_3 &\leq \tau_1 = \gamma, \\
    \lambda_0 + 3 \lambda_1 + 3 \lambda_2 + \lambda_3 &\leq \tau_0 = 1.
\end{align}
Motivated by the feasible solution of $\lambda_\star$ for the constraints in~\eqref{eq:lambdaconstraintsonK=3}, the server finds multicast messages to serve all users, as shown later in the delivery phase.
\label{eq:lambdaconstraintsonK=3}
\end{subequations}

\paragraph*{Delivery Phase}
For the sake of simplicity, we assume that user $k$ demands $F_k$, for every $k \in [3]$. For arbitrary demands and memory ratio $\gamma$, please refer to the proof of achievability in Section~\ref{sec:proofofHT1}.

\paragraph{Case $0\leq\gamma\leq1/3$}
A feasible solution for the constraints in~\eqref{eq:lambdaconstraintsonK=3} is:
\begin{align*}
    \lambda_3 = \lambda_2 = 0, \lambda_1 = \gamma, \lambda_0 = 1-3\gamma.
\end{align*}
Equivalently, each file $F_n$ is partitioned into four disjoint and linearly independent coded subfiles: $\{C_{n,\emptyset}, C_{n,1}, C_{n,2}, C_{n,3}\}$.
The server then sends the messages as follows:
\begin{align*}
    \{C_{n,\emptyset}: n \in [3]\} \cup \{C_{2,1} + C_{1,2}, C_{3,1} + C_{1,3}, C_{3,2} + C_{2,3}\}.
\end{align*}
The load is thus $3 \lambda_0 + \lambda_1 = 3-6\gamma$.
For decoding correctness, from the perspective of user~1, it can compute $C_{1,2}$ and $C_{1,3}$ from the first two messages, as it caches $C_{2,1}$ and $C_{3,1}$, respectively. Together with $C_{1, \emptyset}$, and $C_{1,1}$ cached by itself, user~1 can reconstruct its desired file $F_1$. All other users are handled similarly.

\paragraph{Case $1/3\leq\gamma\leq1/2$}
A feasible solution for the constraints in~\eqref{eq:lambdaconstraintsonK=3} is:
\begin{align*}
    \lambda_1 = 1/3, \ \lambda_0 = \lambda_2 = \lambda_3 = 0.
\end{align*}
Let $\eta = (1-\gamma)/2 \leq \lambda_1$, and let $C'_{n,k}$ be the first $\eta\Bsf$ symbols of $C_{n,k}$. The server sends the messages as follows:
\begin{align*}
    \{C'_{2,1} + C'_{1,2}, C'_{3,1} + C'_{1,3}, C'_{3,2} + C'_{2,3}\}.
\end{align*}  
The load is thus $3\eta = 3(1-\gamma)/2$.
For decoding correctness, from the perspective of user~1, it can compute $C'_{1,2}$ and $C'_{1,3}$ from the first two messages, as it caches $C'_{2,1}$ and $C'_{3,1}$, respectively. Together with $C_{1,1}$ cached by itself, user~1 now has 
\begin{align*}
    \Bsf(\gamma + 2\eta) = \Bsf (\gamma + 1 - \gamma) = \Bsf
\end{align*}
linearly independent coded symbols and thus can reconstruct its desired file $F_1$. All other users are handled similarly.

\paragraph{Case $1/2\leq\gamma\leq2/3$}
A feasible solution for the constraints in~\eqref{eq:lambdaconstraintsonK=3} is:
\begin{align*}
    \lambda_2 = 2\gamma-1, \lambda_1 = 2-3\gamma, \ \lambda_3 = \lambda_0 = 0.
\end{align*}
Let $C'_{n, k}$ be the first $\eta_1\Bsf$ symbols of $C_{n,k}$, where 
\begin{align*}
    2 \eta_1 + \lambda_2 = 1-\gamma.
\end{align*}
The server sends the messages as follows:
\begin{align*}
    \begin{Bmatrix}
        C'_{2,1} + C'_{1,2}, C'_{3,1} + C'_{1,3}, C'_{3,2} + C'_{2,3} \\
        C_{1,23} + C_{2,13} + C_{3,12}
    \end{Bmatrix}.
\end{align*}  
The load is thus $3\eta_1 + \lambda_2 = 2 - 5\gamma/2$.
For decoding correctness, from the perspective of user~1, it can compute $C'_{1,2}, C'_{1,3}$, and $C_{1,23}$ since it already caches the other coded subfiles. Together with its cache content, user~1 now has 
\begin{align*}
    \Bsf(\gamma + 2\eta_1 + \lambda_2) = \Bsf
\end{align*}
linearly independent coded symbols and can thus reconstruct its desired file $F_1$. All other users are handled similarly.

\paragraph{Case $2/3\leq\gamma\leq1$}
A feasible solution for the constraints in~\eqref{eq:lambdaconstraintsonK=3} is
$
    \lambda_3 = 3\gamma-2, \lambda_2 = 1-\gamma, \ \lambda_1 = \lambda_0 = 0.
$
The server sends the message as follows:
$
    C_{1,23} + C_{2,13} + C_{3,12}.
$
The load is thus $\lambda_2 = 1-\gamma$.
For decoding correctness, from the perspective of user~1, it can compute $C_{1,23}$ since it caches $C_{2,13}$ and $C_{3,12}$. Together with its cache content, user~1 now has $\Bsf$
linearly independent coded symbols and can thus reconstruct its desired file $F_1$. All other users are handled similarly.

\paragraph*{Performance}
Fig.~\ref{fig: hotplug (3,3)} summarizes the memory-load performance achieved by the $1$-LinP scheme. Later, in Section~\ref{sec:converse}, we prove that it is exactly optimal when the placement is $1$-LinP. In contrast, we also show the exact memory-load performance when the placement is uncoded, as stated in Theorem~\ref{thm:extensionYMAdecentralized}. Note that the curve of the $1$-LinP scheme is not convex, as memory sharing is impossible (see Remark~\ref{rem:memorysharingimpossible}).

\begin{figure}
   \centering 
    \begin{tikzpicture}
    \begin{axis}[
        xmin= 0, xmax=1,
        ymin= 0, ymax=3,
        xtick distance=0.2,
        ytick distance=0.5, 
        legend entries={ 
            \small $1$-LinP in~\eqref{eq:lpachievability},
            \small Uncoded in~\eqref{eq:performanceDecentralized}
        },
        height=5.5cm,
        width=8cm,
        ylabel near ticks,
        grid=major,
        grid style=dashed,
        legend pos=north east,
        xlabel={\small Memory ratio $\gamma$},
        ylabel={\small Worst-case load $\Delta$}
    ]
    \addplot[color=red!70, mark=*, semithick] coordinates {
        (0,3) (0.333, 1) (0.5, 0.75) (0.666, 0.333) (1, 0)
    };
    \addplot[color=gray, semithick] table [x=x, y=y] {data/decentralized3.data};
    \end{axis}
    \end{tikzpicture}

    \caption{\small Memory-load tradeoffs for the decentralized coded caching system with $\Nsf=3$ files and $\Ksf=3$ active users.}
    \label{fig: hotplug (3,3)}
\vspace*{-5mm}
\end{figure}
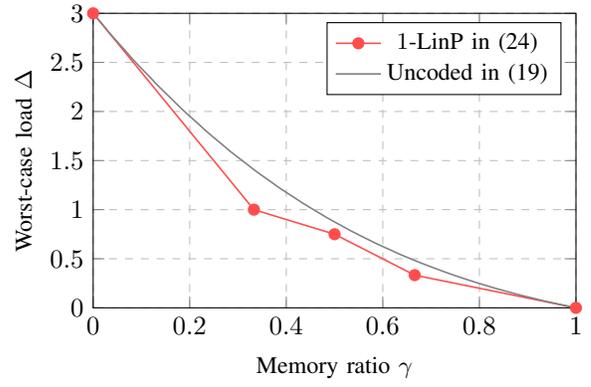

\begin{figure*}
    \centering
    \begin{subfigure}[b]{0.45\textwidth}
        \centering 
        \scalebox{0.8}{
        \begin{tikzpicture}[every pin/.style={fill=white}]
        \begin{axis}[
            xmin=0, xmax=1,
            ymin=0, ymax=3,
            legend entries={
                \small $1$-LinP~\eqref{eq:lpachievability},
                \small Uncoded~\eqref{eq:performanceDecentralized},
                \small MDS~\eqref{eq:mdsdecendlp},
                \small Converse~\eqref{eq:mainconverse}\cite{yu2018characterizing}
            },
            height=8cm,
            width=10cm,
            grid=major,
            ylabel near ticks,
            grid style=dashed,
            legend pos=north east,
            xlabel={\small Memory ratio $\gamma$},
            ylabel={\small Achievable load $\Rsf$}
        ]
        \addplot[color=red!70, thick]  table [x=x, y=y1] {data/example36.data};
        \addplot[color=blue!80, thick]  table [x=x, y=y2] {data/example36.data};
        \addplot[color=magenta!70, thick]  table [x=x, y=y3] {data/example36.data};
        \addplot[color=black, densely dashdotted, thick]  table [x=x, y=y4] {data/example36.data};
        \coordinate (spypoint) at (axis cs:0.8, 0.2);
        \end{axis}
        \node[pin={[pin edge={->,black!70,semithick},pin distance=0.8cm]90:{%
        \begin{tikzpicture}[baseline,trim axis left,trim axis right,scale=1.25]
            \begin{axis}[
                    tiny,ticks=none,
                    xmin=0.72,xmax=0.9,
                    ymin=0.1,ymax=0.4,
                    yticklabels={,,},
                    xticklabels={,,}
                ]
                \addplot[color=red!70, thick]  table [x=x, y=y1] {data/example36.data};
                \addplot[color=blue!80, thick]  table [x=x, y=y2] {data/example36.data};
                \addplot[color=magenta!70, thick]  table [x=x, y=y3] {data/example36.data};
                \addplot[color=black, densely dashdotted, thick]  table [x=x, y=y4] {data/example36.data};
            \end{axis}
        \end{tikzpicture}%
        }},draw,circle,minimum size=0.75cm] at (spypoint) {};
        \end{tikzpicture}
        }
        \caption{\small Case $(\Ksf,\Nsf) = (6, 3)$.}
        \label{fig:example63}
    \end{subfigure}
    \begin{subfigure}[b]{0.45\textwidth}
        \centering 
        \scalebox{0.8}{
        \begin{tikzpicture}[every pin/.style={fill=white}]
        \begin{axis}[
            xmin=0, xmax=1,
            ymin=0, ymax=6,
            legend entries={
                \small $1$-LinP~\eqref{eq:lpachievability},
                \small Uncoded~\eqref{eq:performanceDecentralized},
                \small MDS~\eqref{eq:mdsdecendlp},
                \small Converse~\eqref{eq:mainconverse}\cite{yu2018characterizing}
            },
            height=8cm,
            width=10cm,
            grid=major,
            grid style=dashed,
            legend pos=north east,
            ylabel near ticks,
            xlabel={\small Memory ratio $\gamma$},
            ylabel={\small Achievable load $\Rsf$}
        ]
        \addplot[color=red!70, thick]  table [x=x, y=y1] {data/example66.data};
        \addplot[color=blue!80, thick]  table [x=x, y=y2] {data/example66.data};
        \addplot[color=magenta!70, thick]  table [x=x, y=y3] {data/example66.data};
        \addplot[color=black, densely dashdotted, thick]  table [x=x, y=y4] {data/example66.data};
        \coordinate (spypoint) at (axis cs:0.81, 0.3);
        \end{axis}
        \node[pin={[pin edge={->,black!70,semithick},pin distance=0.5cm]90:{%
        \begin{tikzpicture}[baseline,trim axis left,trim axis right,scale=1.2]
            \begin{axis}[
                    tiny,ticks=none,
                    xmin=0.72,xmax=0.9,
                    ymin=0.1,ymax=0.4,
                    yticklabels={,,},
                    xticklabels={,,}
                ]
                \addplot[color=red!70, thick]  table [x=x, y=y1] {data/example66.data};
                \addplot[color=blue!80, thick]  table [x=x, y=y2] {data/example66.data};
                \addplot[color=magenta!70, thick]  table [x=x, y=y3] {data/example66.data};
                \addplot[color=black, densely dashdotted, thick]  table [x=x, y=y4] {data/example66.data};
            \end{axis}
        \end{tikzpicture}%
        }},draw,circle,minimum size=0.75cm] at (spypoint) {};
        \end{tikzpicture}
        }
        \caption{\small Case $(\Ksf,\Nsf) = (6, 6)$.} 
        \label{fig:example66}
    \end{subfigure}
    \caption{\small Memory-load tradeoffs for the decentralized system, the converse is derived from~\eqref{eq:mainconverse} and bounds for general placement in~\cite{yu2018characterizing}.}
    \label{fig:numericalexamples} \vspace*{-3mm}
\end{figure*}
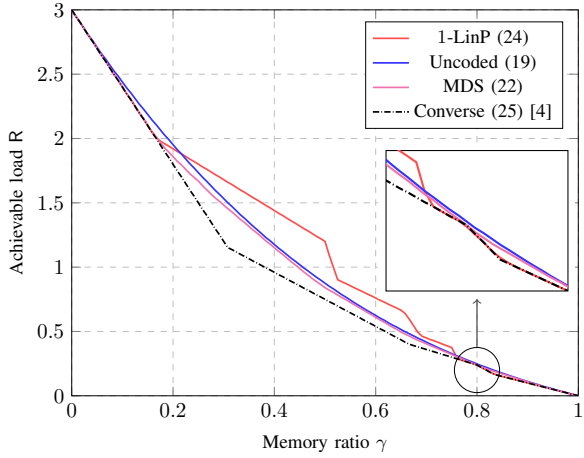
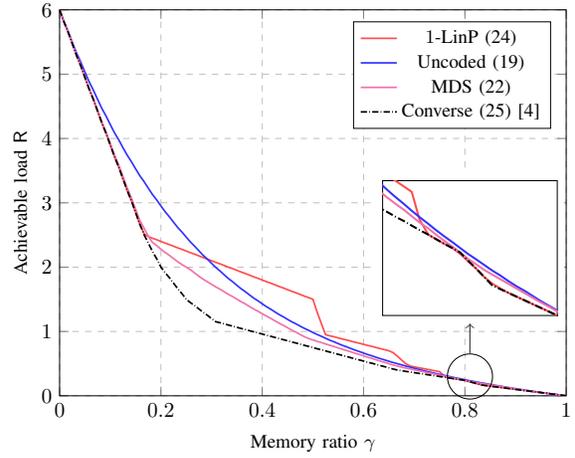

\subsection{Achievability Imposed by Linear Program}
\label{sec:achievabilitylinearprogram}
We present the achievability of the decentralized system under $1$-LinP. In the rest of the paper, we refer to the achievable scheme presented in Theorem~\ref{thm:HT1} as the $1$-LinP scheme.

\begin{thm}%
    \label{thm:HT1}
    For a decentralized coded caching system with $1$-LinP, when $\Nsf$ files and $\Ksf$ active users are present, let $r := \min(\Nsf, \Ksf)$, the expected worst-case load is upper bounded by
    \begin{subequations}
    \begin{align}
        \Delta^{(\mathfrak{L}_1)}(\gamma) \leq \Rsf^{(\mathfrak{L}_1)}(\gamma) = \min_{\lambda_\star, \eta_\star} \sum_{j = 0}^{\Ksf-1} \eta_j %
        c(j), \label{eq:lpobjectivefunction}
    \end{align}
    where $c(t) := \binom{\Ksf}{t+1} - \binom{\Ksf-r}{t+1}$, and the optimization variables are $\{\lambda_j, \eta_j: j \in [0: \Ksf]\}$. The constraints are as follows,
    \begin{align}
        \sum_{j=s}^{\Ksf} \binom{\Ksf-s}{j-s}\lambda_j &\leq \left[1 - s(1-\gamma)\right]^+, \label{eq:lpconstraintsonlambdas}\\
        \sum_{j=0}^{\Ksf-1} \binom{\Ksf-1}{j} \eta_j &= 1-\gamma, \label{eq:decodingcorrectness}\\
        0 \leq \eta_s &\leq \lambda_s, \label{eq:upperboundmessage}
    \end{align}
    for all $s \in [0:\Ksf]$. %
    \label{eq:lpachievability}
    \end{subequations}
\end{thm}
    Similar to the decentralized scheme with uncoded placement in~\cite{maddah2014decentralized,yu2017exact}, Theorem~\ref{thm:HT1} first identifies which coded subfiles (measured by $\lambda_\star$) are computed among certain users only to facilitate multicast opportunities and find messages (measured by $\eta_\star$) so users can exchange these coded subfiles. 
    The essence of constraint~\eqref{eq:decodingcorrectness} is decoding correctness to ensure that users know exactly $\Bsf$ independent coded symbols to reconstruct their desired scalar linear function of files.
    When $\Nsf \geq 2$ and $\Ksf \geq 2$, the analytical solutions of~\eqref{eq:lpachievability} is attained when $\gamma$ satisfies certain conditions, as shown in Table~\ref{tab:generalsolutions}.
    The proof is provided in Section~\ref{sec:proofofHT1}. \hfill$\square$
\begin{rem}
    Optimization is a powerful framework and has been used to derive bounds for caching systems with various settings, 
    including Femtocaching~\cite{shanmugam2013femtocaching}, MAN system~\cite{tian2018symmetry},heterogeneous coded caching~\cite{daniel2019optimization}, and linear computational broadcasting channel (LCBC)~\cite{yao2024capacity,ma2025achievablearxiv}. 
    Theorem~\ref{thm:HT1} is closely related to recent works~\cite{yao2024capacity,ma2025achievablearxiv} in LCBC, in which they propose a subspace decomposition on matrices and derive an achievable scheme via an elaborated linear program.
\end{rem}

\subsection{Numerical Example}
We provide two additional examples to illustrate the performance of $1$-LinP scheme in Fig.~\ref{fig:numericalexamples}, where
Fig.~\ref{fig:example63} shows the memory-load tradeoff $(\Ksf, \Nsf) = (6, 3)$, and Fig.~\ref{fig:example66} is for $(\Ksf, \Nsf) = (6, 6)$.
The converse is the maximal value between our proposed converse~\eqref{eq:mainconverse} under the constraint of $1$-LinP placement, and converse bounds for general placement in~\cite[Theorem~2, Theorem~4]{yu2018characterizing}.
The $1$-LinP scheme reaches optimality when the memory ratio is small or large, as shown in Table~\ref{tab:generalsolutions}. 
We note that the load performance of the MDS scheme in~\eqref{eq:mdsdecendlp} is not always achievable if the cache content is immutable after the placement phase (see Remark~\ref{rem:mds}).

\begin{table*}
\renewcommand{\arraystretch}{2}
\centering
\caption{The memory-load tradeoff of $1$-LinP scheme in Theorem~\ref{thm:HT1} with at least $\Nsf=2$ files and $\Ksf=2$ users.\label{tab:generalsolutions}}
\begin{tabular}{ | c || c | c | }
\hline
Memory ratio $\gamma$ & Achievability Load $\Rsf^{(\mathfrak{L}_1)}(\gamma)$ & Match Converse~\eqref{eq:mainconverse}? \\ 
\hline
$[0, \ 1\big/\Ksf]$ & $r - \gamma r (r+1)\big/2$ & Yes \\ 
\hline
$[1\big/\Ksf, \ 1\big/2]$ & $(1-\gamma) r (2\Ksf-r-1)\big/(2\Ksf-2)$ & Only when $\Ksf=3$ \\ 
\hline
$t\big/(t+1), \forall t \in \{1,\ldots,\Ksf-3\}$ & $c(t)\big/\left((t+1) \binom{\Ksf-1}{t}\right)$ & No when $\Ksf \geq 4$ \\ 
\hline
$[(\Ksf-2)\big/(\Ksf-1), \ (\Ksf-1)\big/\Ksf]$ & $2 - \gamma(2\Ksf-1)\big/(\Ksf-1)$ & Yes \\ 
\hline
$[(\Ksf-1)\big/\Ksf, \  1]$ & $ 1 - \gamma$ & Yes \\ 
\hline
\end{tabular}
\renewcommand{\arraystretch}{1}
\end{table*}

\section{Converse with $1$-LinP Placement}
\label{sec:converse}
We present the lower bound on the expected worst-case load of the decentralized system under $1$-LinP. 
\begin{thm}%
    \label{thm:converse}
    For a decentralized coded caching system, let $r = \min(\Nsf,\Ksf)$, the expected worst-case load with $1$-LinP is lower-bounded by
    \begin{subequations}
    \begin{align}
        \Delta^{(\mathfrak{L}_1)}(\gamma) &\geq \Delta_1 = \sum_{t=1}^{r} \left[1 - t\gamma\right]^+. \label{eq:conversedelta1}
    \end{align}
    If $\Ksf \geq 3$ and $\Nsf \geq 2$, then
    \begin{align}
        \Delta^{(\mathfrak{L}_1)}(\gamma) \geq \Delta_2 &=  r(1-\gamma) - \frac{\tau_{\Ksf-1}-\tau_{\Ksf}}{\Ksf-1} - \sum_{j=3}^{r} \left(\rho_j-\gamma \right)
    \notag \\ & - \sum_{j=3}^{\Ksf}\left(\frac{\min(\gamma+\tau_{j-2},1-\gamma)}{\Ksf-1} \right), \label{eq:conversedelta2}
    \end{align}
    where $\tau_s = [1-s(1-\gamma)]^+$ and $\ \rho_s = \min(s\gamma, 1).$ %
    \label{eq:mainconverse}   \hfill$\square$
    \end{subequations}
\end{thm}

The first converse bound in Theorem~\ref{thm:converse}, $\Delta_1$, is the genie-aided bound.
The second converse bound, $\Delta_2$, is inspired by related work in~\cite{yao2024capacity}. It leverages functional submodularity and the rank of unions and intersections between any two pairwise matrices from the same column space.
For the achievability, as shown in Table~\ref{tab:generalsolutions},
the $1$-LinP scheme is exactly optimal when $\Ksf \leq 3$, and otherwise in the small or large memory regime when $\Ksf \geq 4$.
The detailed proof is provided in Section~\ref{sec:proofofconverse}. 

We now prove the optimality claims in Table~\ref{tab:generalsolutions}.
\paragraph{Case $\gamma \in [0, 1/\Ksf]$}
\eqref{eq:conversedelta1} yields
\begin{align*}
    \Delta_1 = \sum_{t=1}^{r} (1 - t\gamma) = r - \frac{r(r+1)}{2}\gamma,
\end{align*}
which matches the load of the $1$-LinP scheme.
\paragraph{Case $\gamma \in [1/3, 1/2]$ when $\Ksf = 3$}
\eqref{eq:conversedelta2} yields $
    \Delta_2 = 3(1-\gamma)/2,
$
since $\rho_3 = 1$ and $\tau_s = 0$ for all $s \geq 2$.
Table~\ref{tab:generalsolutions} shows,
\begin{align*}
    \Delta^{(\mathfrak{L}_1)}(\gamma) \leq \frac{r(5-r)}{4}(1-\gamma) = \Delta_2.
\end{align*}
\paragraph{Case $\gamma \in [(\Ksf-2)/(\Ksf-1), (\Ksf-1)/\Ksf]$} \eqref{eq:conversedelta2} yields 
\begin{align*}
    \Delta_2 &= \frac{\Ksf(1-\gamma)}{\Ksf-1} - \frac{1-(\Ksf-1)(1-\gamma)}{\Ksf-1} \\
    &= 2 - \frac{\gamma(2\Ksf-1)}{\Ksf-1},
\end{align*}
since $\tau_\Ksf = 0$ and $\rho_s = 1$ for all $s \geq 3$. This matches the load of the $1$-LinP scheme.

\paragraph{Case $\gamma \in [(\Ksf-1)/\Ksf, 1]$}
\eqref{eq:conversedelta1} yields
$\Delta_1 =  1 - \gamma,$
which matches the load of the $1$-LinP scheme.

\section{Proof of Achievability in Theorem~\ref{thm:HT1}}
\label{sec:proofofHT1}
We introduce the $1$-LinP scheme and derive the linear program~\eqref{eq:lpachievability} in Theorem~\ref{thm:HT1}.

\paragraph*{Placement Phase} 
Fix $\gamma \in [0,1]$, for every $k \in \ZZ^+$ and $n \in [\Nsf]$, the server constructs a matrix $\Em_{k,n} \in \FF_\qsf^{\Bsf \times \Bsf \gamma }$, each element of $\Em_{k,n}$ is chosen i.i.d. uniformly in $\FF_\qsf$. Then, for every user $k$, the server populates the cache content $Z_k$ as follows,
\begin{align}
    Z_k = \{\Em_{k,n}^T F_n: n \in [\Nsf]\}.
\end{align}

\paragraph*{Delivery Phase}
Given $n \in [\Nsf]$ and $\Sc \subseteq [\Ksf]$, let $\Em_{\Sc,n} = \cap_{k \in \Sc} \langle \Em_{k,n}\rangle$.
Similar to the decentralized scheme in Subsection~\ref{sec:DecentralizedPlacement+hotplug}, 
$\Cm_{n, \Sc} := \Vm_{\Sc, n}^T F_n$ is the subfile of $F_n$ which can be computed by users $\Sc$ \textbf{only}, where $\Vm_{\Sc, n} \in \FF_\qsf^{\Bsf \times \rk(\Vm_{\Sc, n})} \subseteq \langle \Em_{\Tc, n} \rangle$ for all $\Tc \subseteq \Sc \subseteq [\Ksf]$.  %
Intuitively, $\Vm_{\Sc,n}$ is the basis that forms the subspace $\Em_{\Sc,n}$ only.
By definition, 
for every subset $\Tc \subseteq [\Ksf]$, $\Em_{\Tc,n}$ covers all subspaces $\Vm_{\Sc, n}$ where $\Tc \subseteq \Sc$, 
\begin{align}
    \rk(\Em_{\Tc, n}) \geq \rk(\bigcup_{\Sc \subseteq [\Ksf], \Tc \subseteq \Sc} \Vm_{\Sc, n})
    \stackrel{(a)}{=} \sum_{\Sc \subseteq [\Ksf], \Tc \subseteq \Sc} \rk(\Vm_{\Sc, n}),
    \label{eq:basiscoveredbysubset}
\end{align}
where $(a)$ follows from the fact that every $\Vm_{\Sc, n}$ is linearly independent, 
Due to file and user symmetry, the rank of $\Vm_{\Sc, n}$ and $\Em_{\Sc,n}$ depends only on $s = |\Sc|$. Thus, let $\lambda_s := \rk(\Vm_{\Sc, n})/\Bsf$ and $\tau_s := \rk(\Em_{\Sc, n})/\Bsf$, \eqref{eq:basiscoveredbysubset} is thus rewritten as,
\begin{align}
    \tau_s \geq \sum_{t=s}^{\Ksf} \binom{\Ksf-s}{t-s} \lambda_t, \ \forall s \in [\Ksf].
    \label{eq:constraintsonlambdas}
\end{align}
By convention, we let $\tau_0 = \rk(\Id_\Bsf)/\Bsf = 1$ and $\Vm_{\emptyset, n} := \langle\Id_\Bsf\rangle \setminus \Em_{([\Ksf]),n}$. That is, $\Cm_{\emptyset, n} := \Vm_{\emptyset, n}^T F_n$ are the symbols that are not covered by any user. 
\eqref{eq:constraintsonlambdas} is exactly the constraint~\eqref{eq:lpconstraintsonlambdas} where $s \in [0:\Ksf]$.

Similar to~\eqref{eq:SLFRblock}, given the demand matrix $\Dm \in \FF_\qsf^{\Ksf \times \Nsf}$, with a slight abuse of notation, we define
\begin{align}
    \Bm_{k, \Sc} := \sum_{n \in [\Nsf]} d_{k,n} \Cm_{n, \Sc} \in \FF_\qsf^{\Bsf\lambda_{|\Sc|}}, \ \forall \Sc \subseteq [\Ksf], \label{eq:slfrcodedblock}
\end{align}
where $d_{k,n}$ is the element from the demand matrix $\Dm$.
For every $j \in [0:\Ksf]$, let $\eta_j \leq \lambda_j$. We truncate the coded subfile $\Cm_{n,\Sc}$, let the first $\eta_j\Bsf$ symbols of $\Cm_{n,\Sc}$ be $\Cm'_{n,\Sc}$, and define
\begin{align}
    \Bm'_{k, \Sc} := \sum_{n \in [\Nsf]} d_{k,n} \Cm'_{n, \Sc} \in \FF_\qsf^{\Bsf\eta_{|\Sc|}}, \ \forall \Sc \subseteq [\Ksf].
\end{align}

For every $\Sc \subseteq [\Ksf]$ and $s := |\Sc| \geq 1$, the server constructs a multicast (or unicast when $s=1$) message for the users in $\Sc$,
\begin{align}
    X_\Sc = \sum_{k \in \Sc} \alpha_{\Sc, k} \Bm'_{k, \Sc \setminus \{k\}} \in \FF_\qsf^{\eta_{s-1}}, \label{eq:codedmultiSusers}
\end{align}
where $\alpha_\star \in \{1, -1\}$ are the encoding coefficients selected as shown in~\cite{ma2021general}.

The server identifies the leader set $\Lc \subseteq [\Ksf]$, where $\rk(\Dm[\Lc]) = \rk(\Dm)$, and then sends the message $X$,
\begin{align}
    X = \{X_\Sc: \Sc \subseteq [\Ksf], |\Sc| \geq 1, \Sc \cap \Lc \neq \emptyset\}.
\end{align}
The load is then $$\sum_{j=0}^{\Ksf-1} \left(\binom{\Ksf}{j+1}-\binom{\Ksf-|\Lc|}{j+1}\right) \eta_j,$$ 
which is exactly the objective function~\eqref{eq:lpobjectivefunction} when $|\Lc| = \min(\Ksf,\Nsf)$.

Any user $j \in \Sc$ can decode the truncated coded subfile $\Bm'_{j, \Sc \setminus \{j\}}$ since, from~\eqref{eq:codedmultiSusers},
\begin{align}
    X_\Sc &= \alpha_{\Sc, j} \Bm'_{j, \Sc \setminus \{j\}} + \sum_{k \in \Sc \setminus \{j\}} \alpha_{\Sc, k} \Bm'_{k, \Sc \setminus \{k\}} \notag \\
    &= \alpha_{\Sc, j} \Bm'_{j, \Sc \setminus \{j\}} + \sum_{k \in \Sc \setminus \{j\}} \alpha_{\Sc, k} \sum_{n \in [\Nsf]} d_{k,n} \Cm'_{n, \Sc},
\end{align}
where the last term $\Cm'_{n, \Sc} \subseteq \Cm_{n, \Sc}$ is known by user $j$, who can thus extract $\Bm'_{j, \Sc \setminus \{j\}}$ locally.
Each user $j$ is able to compute all $$\{\Bm'_{j, \Sc}: \Sc \subseteq [\Ksf]\setminus\{j\} \},$$ %
Together with the $\gamma\Bsf$ symbols cached by user~$j$, user~$j$ can decode its desired scalar linear function $B_j$ if
$$
\sum_{j=0}^{\Ksf-1} \binom{\Ksf-1}{j} \eta_j + \gamma = 1,
$$
which is exactly constraint~\eqref{eq:decodingcorrectness}.

We now give the feasible solutions to attain the results in Table~\ref{tab:generalsolutions}, note that 
$$c(1) = \binom{\Ksf}{2} - \binom{\Ksf-r}{2} = \Ksf r - \frac{r(r+1)}{2}.$$
\paragraph{Case $0 \leq \gamma \leq 1/\Ksf$}
Equivalently, we have %
\begin{align*}
    \eta_1 = \lambda_1 = \tau_1 = \gamma , \
    \eta_0 = \lambda_0 = \tau_0 = 1 - \Ksf \tau_1 = 1-\Ksf\gamma,
\end{align*}
and other $\eta_\star=\lambda_\star=0$.
The load is $ 
    r(1-\Ksf\gamma) + \gamma c(1) = r - \gamma r(r+1)/2
$.

\paragraph{Case $1/\Ksf \leq \gamma \leq 1/2$ where $\Ksf \geq 3$}
Equivalently, we have $\Ksf\gamma \geq 1 \geq 2\gamma$, which implies 
$
     \tau_1 = \gamma, \ \tau_2 = 0.
$
The constraints~\eqref{eq:constraintsonlambdas} then becomes $\lambda_1 \leq \tau_1$, $\Ksf \lambda_1 + \lambda_0 \leq \tau_0 = 1$. 
A feasible solution is $\lambda_1 = 1/\Ksf$ and $\eta_1 = (1-\gamma)/(\Ksf-1)$,
and other $\eta_\star = \lambda_\star = 0$.
The load is $(1-\gamma)c(1)/(\Ksf-1) = (1-\gamma) r (2\Ksf-r-1)/(2\Ksf-2)$.

\paragraph{Case $\gamma = t/(t+1)$ for $t \in [\Ksf-1]$}
Equivalently we have
$\tau_{s} = 0$ for all $s \geq t+1$, and $\tau_s = 1-s/(t+1)$ for all $s \leq t$. The optimal solution is: 
\begin{align*}
    \lambda_{t} = 1/\binom{\Ksf}{t}, \eta_t = 1/[(t+1)\binom{\Ksf-1}{t}],
\end{align*}
and other $\lambda_\star = \eta_\star = 0$.
Note that,
$$\frac{\eta_t}{\lambda_t} = \frac{1}{t+1} \frac{\Ksf}{\Ksf-t} \stackrel{(a)}{\leq} \frac{1}{t+1} \frac{t+1}{t+1-t} = 1,$$
where $(a)$ follows from $\Ksf \geq t+1$. The load is $c(t)\eta_{t} = c(t)/\left((t+1)\binom{\Ksf-1}{t}\right)$.
\paragraph{Case $(\Ksf-2)/(\Ksf-1) \leq \gamma \leq (\Ksf-1)/\Ksf$}
Equivalently we have
\begin{align*}
\tau_{\Ksf-2} &= 1 - (\Ksf-2) (1-\gamma) \in [1/(\Ksf-1), 2/\Ksf]. \\
\tau_{\Ksf-1} &= 1 - (\Ksf-1) (1-\gamma) \in [0, 1/\Ksf], 
\end{align*}
A feasible solution is $\eta_{\Ksf-1} = \lambda_{\Ksf-1} = \tau_{\Ksf-1}$, $\lambda_{\Ksf-2} = (1-\Ksf\lambda_{\Ksf-1})/\binom{\Ksf}{2}$, and $\eta_{\Ksf-2} = (1-\gamma-\eta_{\Ksf-1})/(\Ksf-1)$.
The load is $
\Ksf \eta_{\Ksf-2} + \eta_{\Ksf-1} = 2 - \gamma(2\Ksf-1)/(\Ksf-1).
$

\paragraph{Case $(\Ksf-1)/\Ksf \leq \gamma \leq 1$}
Equivalently we have
\begin{align*}
\tau_{\Ksf-1} &= 1 - (\Ksf-1) (1-\gamma) \in [0, 1/\Ksf], \\
\tau_\Ksf &= 1 - \Ksf (1-\gamma) \in [0, 1].
\end{align*}
A feasible solution is $\eta_\Ksf=\lambda_\Ksf = \tau_\Ksf$ and $\eta_{\Ksf-1}=\lambda_{\Ksf-1} = (1-\lambda_\Ksf)/\Ksf = 1-\gamma $, with other $\eta_\star=\lambda_\star = 0$.
The load is $\eta_{\Ksf-1} = 1-\gamma.$

\section{Proof of Converse in Theorem~\ref{thm:converse}}
\label{sec:proofofconverse}
Given two matrices $\Mm_1, \Mm_2$ from the same row space, we define the conditional rank as
\begin{align*}
    \rk(\Mm_1 | \Mm_2) &:= \rk([\Mm_1; \Mm_2]) - \rk(\Mm_2) \\
    &= \rk(\Mm_1) - \rk(\Mm_1 \cap \Mm_2).
\end{align*}
\subsection{$\Delta^{(\mathfrak{L}_1)} \geq \Delta_1$}
Without loss of generality, let $r := \min(\Ksf,\Nsf)$. For every $k \in [\Ksf]$, we assume $d_k = \min(k, \Nsf)$; that is, the leader users are the first $r$ users, $\Lc = [r]$. Then, by the genie-aided method in~\cite{yu2017exact,wan2020index}, we have
\begin{align}
    H(X) &\geq \sum_{j=1}^r H(F_j | F_{j-1}, \ldots, F_1, Z_j, \ldots, Z_1) 
    \label{eq:conversegenieaidefirststage}
\end{align}
Note that, given $s \in [\Nsf]$ and $t \in [\Ksf]$, let
\begin{align*}
    g(s, t) &:= H(F_s, \ldots, F_1, Z_t, \ldots, Z_1) \\
    &=H(F_s, \ldots, F_1) + H(Z_t,\ldots,Z_1 | F_s, \ldots, F_1) \\
    &=s \ \rk(\Id_\Bsf) + \sum_{n=s+1}^{\Nsf}\rk([\Em_{1,n},\ldots,\Em_{t,n}]) \\
    &= s \Bsf + (\Nsf-s) \min(1, t\gamma) \Bsf,
\end{align*}
\eqref{eq:conversegenieaidefirststage} is equivalent to,
\begin{align}
    H(X) &\geq \sum_{j=1}^r g(j, j) - g(j-1, j) \notag \\
    &= \sum_{j=1}^r (\Bsf - \Bsf \min(1, j \gamma)) \notag \\
    &= \Bsf \sum_{j=1}^r \left[1 - j \gamma \right]^+  
    = \Delta_1.
    \label{eq:conversegenieaidefinalstage}
\end{align}

\subsection{$\Delta^{(\mathfrak{L}_1)} \geq \Delta_2$}
Recall that the concatenation of all files is denoted by $F = [F_1, \ldots, F_n] \in \FF_\qsf^{\Nsf\Bsf}$; by convention, we define $\Fm_n \in \FF_\qsf^{\Nsf\Bsf \times \Bsf}$ such that $F_n = \Fm_n^T F$. 
Let $r := \min(\Ksf,\Nsf)$. For every $k \in [\Ksf]$, we assume $d_k = \min(k, \Nsf)$; that is, the leader users are the first $r$ users, $\Lc = [r]$.
Recall that the cache content is populated as $Z_k = \Em_k^T F$. We let 
$U_k = [F_k, Z_k]$, and $\Um_k = [\Fm_k, \Em_k]$ for every $k \in [r]$, and
\begin{align*}
    \Um_\Tc = \bigcap_{k \in \Tc} \langle \Um_k \rangle, \ \Um_{(\Tc)} = \bigcup_{k \in \Tc} \langle \Um_k \rangle, \ \text{where }\Tc \subseteq [\Ksf].
\end{align*}
Furthermore, given $k \in [\Ksf]$ and $\Tc \subseteq [\Ksf]\setminus\{k\}$, we let $\Um_{k(\Tc)} := \langle \Um_k \rangle \cap \Um_{(\Tc)}$.
Recall that for some $\Sc \subseteq [\Ksf]$ and $n \in [\Nsf]$, we denote $\Em_{\Sc,n} := \cap_{k \in \Sc} \langle \Em_{k,n}\rangle$ and $\Em_{\emptyset} = \Id_\Bsf$.
\begin{lem}[{Functional submodularity in~\cite[Corollary~1]{yao2024capacity}}]
    \label{lem:submodularitylemma}
    Given any arbitrary matrices $\Mm_1 \in \FF_\qsf^{d \times m_1}, \Mm_2 \in \FF_\qsf^{d \times m_2}$, a random matrix $F \in \FF_\qsf^{d \times L}$, and any random variable $Z$, %
    \begin{multline*}
        H(Z, F^T \Mm_1) + H(Z, F^T \Mm_2) \geq \\ H(Z, F^T [\Mm_1 \cap \Mm_2]) + H(Z, F^T [\Mm_1,\Mm_2])
    \end{multline*}
\end{lem}
We assume that $\Ksf \geq 3$ and $\Nsf \geq 2$. First, note that
\begin{align}
    &\sum_{j \in [\Ksf]} H(X, Z_j) = \sum_{j \in [\Ksf]} H(X, Z_j, F_{d_j}) = \sum_{j \in [\Ksf]} H(X, F^T \Um_j) \notag \\
    &\stackrel{(a)}{\geq} H(X, F^T \Um_{12}) + H(X, F^T \Um_{(12)}) + \sum_{j=3}^{\Ksf} H(X, F^T \Um_{j}) \notag \\
    &\geq \ldots \notag \\
    &\stackrel{(b)}{\geq} H(X, F^T \Um_{([\Ksf])})  + \sum_{j = 2}^\Ksf H(X, F^T \Um_{j(j-1, \ldots, 1)}), \label{eq:intermediate1} 
\end{align}
where $(a)$ and $(b)$ follow from repeated application of Lemma~\ref{lem:submodularitylemma}.
We now examine the term $H(X, F^T \Um_{12})$:
\begin{align}
    &(\Ksf-1) H(X, F^T \Um_{12}) \notag \\
    &= H(X, F^T \Um_{12}) + H(X, F^T \Um_{13}) + \ldots H(X, F^T \Um_{1\Ksf}) \notag \\
    &\stackrel{(a)}{\geq} H(X, F^T \Um_{123}) + H(X, F^T [\Um_{12}, \Um_{13}]) \notag \\&\quad + \sum_{j=4}^\Ksf H(X, F^T \Um_{1j}) \notag \\
    &\stackrel{(b)}{\geq} H(X,  F^T \Um_{1234}) + H(X, F^T [\Um_{12}, \Um_{13}]) \notag \\&\quad + H(X,F^T [\Um_{123}, \Um_{14}]) + \sum_{j=5}^\Ksf H(X, F^T \Um_{1j}) \notag \\
    &\geq \ldots \notag \\
    &\stackrel{(c)}{\geq} H(X, F^T \Um_{[\Ksf]}) + \sum_{j=3}^{\Ksf} H(X, F^T [\Um_{[j-1]}, \Um_{1j}]), \label{eq:intermediate2}
\end{align}
where $(a), (b)$, and $(c)$ follow from repeated application of Lemma~\ref{lem:submodularitylemma}.
Therefore, combining~\eqref{eq:intermediate1} and~\eqref{eq:intermediate2}, we get
\begin{align*}
    &\sum_{j \in [\Ksf]} H(X, Z_j) \\&\geq  H(F^T \Um_{([\Ksf])}) + (\Ksf-1) H(X)  
     + \frac{H(F^T \Um_{[\Ksf]}| X)}{\Ksf-1} \\&\quad + \sum_{j = 3}^\Ksf \frac{H(F^T [\Um_{[j-1]}, \Um_{1j}] | X)}{\Ksf-1}
    \\ &\quad + \sum_{j = 3}^\Ksf H(F^T \Um_{j(j-1, \ldots, 1)} | X).
\end{align*}
On the other hand,
\begin{align*}
    &\sum_{j \in [\Ksf]} H(X, Z_j) = \sum_{j \in [\Ksf]} \left(H(X) + H(Z_j) - I(X; F^T \Em_j)\right) 
    \\ &\leq \Ksf H(X) + \sum_{j\in[\Ksf]} H(Z_j) 
    \\ &\quad - \frac{2}{\Ksf-1} I(X;F^T \Em_1) - \sum_{j = 3}^{\Ksf} I(X;F^T \Em_j),
\end{align*}
since both $I(X;F^T \Em_1)$ and $I(X;F^T \Em_2)$ are non-negative and $\Ksf \geq 3$. Thus, 
\begin{align} 
    &H(X) \geq  H(F^T \Um_{([\Ksf])}) - \sum_{j=1}^{\Ksf} H(Z_j) 
    \notag \\ &\quad + \frac{1}{\Ksf-1}\left(H(F^T \Um_{[\Ksf]} | X) + I(X; F^T \Em_1)\right)
    \notag \\ &\quad + \frac{1}{\Ksf-1}\left(\sum_{j=3}^{\Ksf} H(F^T [\Um_{[j-1]}, \Um_{1j}] | X) + I(X; F^T \Em_1)\right)
    \notag \\ &\quad + \sum_{j=3}^\Ksf \left( H(F^T \Um_{j(j-1,\ldots,1)} | X) +  I(X;F^T \Em_j) \right)
    \notag \\ &\stackrel{(a)}{\geq} \rk(\Um_{([\Ksf])}) - \sum_{j=1}^{\Ksf}\rk(\Em_j) + \frac{\rk(\Um_{[\Ksf]} \cap \Em_1)}{\Ksf-1} 
    \notag \\ &\quad + \sum_{j=3}^{\Ksf} \left(\frac{\rk([\Um_{[j-1]}, \Um_{1j}] \cap \Em_1)}{\Ksf-1}
    + \rk(\Um_{j(j-1,\ldots,1)} \cap \Em_j) \right)
    \notag \\ &\stackrel{(b)}{=} \rk(\Um_{([\Ksf])}) + \frac{\rk(\Um_{[\Ksf]}) + \sum_{j=3}^{\Ksf} \rk([\Um_{[j-1]}, \Um_{1j}])}{\Ksf-1}
    \notag \\ &\quad+ \sum_{j=3}^{\Ksf} \rk(\Um_{j(j-1,\ldots,1)}) - \sum_{j=1}^{\Ksf} \rk(\Em_j) - \frac{\rk(\Um_{[\Ksf]} | \Em_1)}{\Ksf-1} 
    \notag \\ &\quad- \sum_{j=3}^{\Ksf} \frac{\rk([\Um_{[j-1]}, \Um_{1j}] | \Em_1)}{\Ksf-1}
    - \sum_{j=3}^{\Ksf} \rk(\Um_{j(j-1,\ldots,1)} | \Em_j)
    \notag \\ &\stackrel{(c)}{=} \sum_{j=1}^{\Ksf} \rk(\Um_{j} | \Em_j) - \frac{\rk(\Um_{[\Ksf]} | \Em_1)}{\Ksf-1} - \sum_{j=3}^{\Ksf} \rk(\Um_{j(j-1,\ldots,1)} | \Em_j)
    \notag \\ &\quad - \sum_{j=3}^{\Ksf} \frac{\rk([\Um_{[j-1]}, \Um_{1j}] | \Em_1)}{\Ksf-1},
    \label{eq:Delta2eachterm}
\end{align}
where $(a)$ follows from $H(F^T \Um | X) + I(X; F^T \Vm) \geq H(F^T [\Um \cap \Vm]) = \rk(\Um \cap \Vm)$, $(b)$ follows from the definition of conditional rank, and $(c)$ follows from Lemma~\ref{lem:sumoftwomatrix}.
Considering the $1$-LinP, we now compute each term in~\eqref{eq:Delta2eachterm}.
Note that from Section~\ref{sec:proofofHT1}, for every $n \in [\Nsf]$, $\Sc \subseteq [\Ksf]$ and $s = |\Sc|$, %
\begin{align*}
    \tau_s &:= \rk(\Em_{\Sc,n})/\Bsf \aas \left[1 - s(1 -\gamma)\right]^+, \\
    \rho_s &:= \rk(\Em_{(\Sc), n})/\Bsf \aas \min(s \gamma, 1). 
\end{align*}
When $j-1 \geq \Nsf$, 
\begin{align}
    &\rk(\Um_{j(j-1,\ldots,1)}|\Em_j) = \rk(\Um_{j}|\Em_j) = \Bsf(1-\gamma), \\
    &\rk(\Um_{[j]} | \Em_1) = \sum_{n=1}^{\Nsf-1} \rk(\Em_{[j]\setminus\{n\}, n} | \Em_{1,n}) + \rk(\Em_{[\Nsf], \Nsf} | \Em_{1, \Nsf})
    \notag \\ &\quad = \rk(\Em_{[2:j], 1} | \Em_{1,1}) = \Bsf(\tau_{j-1} - \tau_j).
\end{align}
Otherwise, if $j \leq \Nsf$
\begin{align}
    \rk&(\Um_{j(j-1,\ldots,1)} | \Em_j) \notag \\
    &= \sum_{n=1}^{j-1} \rk(\Em_{j,n} | \Em_{j,n}) + \rk(\Em_{(j-1,\ldots,1),j} | \Em_{j,j}) 
    \notag \\ &\quad+ \sum_{n=j+1}^{\Nsf} \rk(\Em_{j(j-1,\ldots,1),n} | \Em_{j, n}) \notag \\
    &= \rk(\Em_{(j-1,\ldots,1),j} | \Em_{j,j}) = \Bsf(\rho_{j}-\gamma). 
    \\ \rk&(\Um_{[j]} | \Em_1) 
    \notag \\ &= \sum_{n=1}^{j} \rk(\Em_{[j]\setminus\{n\}, n} | \Em_{1,n}) + \sum_{n=j+1}^\Nsf \rk(\Em_{[j],n} | \Em_{1,n}) \notag \\
    &=  \rk(\Em_{[2:j], 1} | \Em_{1,1}) = \Bsf (\tau_{j-1} - \tau_{j}).
\end{align}
We now examine $\rk([\Um_{[j-1]}, \Um_{1j}] | \Em_1)$. Similar to the case of $\rk(\Um_{[j]} | \Em_1)$, only the conditional term with respect to $\Em_{1,1}$ is non-zero:
\begin{align}
    \rk&([\Um_{[j-1]}, \Um_{1j}] | \Em_1)
    = \rk([\Em_{[2:j-1], 1}, \Em_{j, 1}] | \Em_{1,1}) 
    \notag \\ &\quad = \Bsf \min(2\gamma + \tau_{j-2}, 1) - \Bsf \gamma \label{eq:Delta2final}
\end{align}
Thus, combining~\eqref{eq:Delta2eachterm}-\eqref{eq:Delta2final}, we have $\Delta_2$ as follows:
\begin{align}
    H(X)/\Bsf &\geq r(1-\gamma) - \frac{\tau_{\Ksf-1}-\tau_{\Ksf}}{\Ksf-1} - \sum_{j=3}^{r}\left(\rho_j-\gamma\right)
    \notag \\ & \ - \sum_{j=3}^{\Ksf}\left(\frac{\min(\gamma+\tau_{j-2},1-\gamma)}{\Ksf-1} \right)
\end{align}

\section{Conclusion}
\label{sec:conclusion}
In this paper, we focus on the decentralized coded caching system with random linear coding placement. Future directions include deriving the achievability and converse bounds for any arbitrary $\msf$-LinP for $\msf \in [2, \Nsf]$, and designing a prefetching scheme $P_\Em$ to achieve the same load performance as a centralized scheme.

\textbf{Acknowledgment:}
This work was supported in part by NSF Award 2312229, and used Delta at NCSA through allocation ELE240014 from the ACCESS program~\cite{boerner2023access}, which is supported by NSF grants 2138259, 2138286, 2138307, 2137603, and 2138296.


\begin{thebibliography}{10}

\bibitem{maddah2014fundamental}
M.~A. Maddah-Ali and U.~Niesen, ``Fundamental limits of caching,'' {\em IEEE
  Transactions on Information Theory}, vol.~60, no.~5, pp.~2856--2867, 2014.

\bibitem{yu2017exact}
Q.~Yu, M.~A. Maddah-Ali, and A.~S. Avestimehr, ``The exact rate-memory tradeoff
  for caching with uncoded prefetching,'' {\em IEEE Transactions on Information
  Theory}, vol.~64, no.~2, pp.~1281--1296, 2017.

\bibitem{wan2020index}
K.~Wan, D.~Tuninetti, and P.~Piantanida, ``An index coding approach to caching
  with uncoded cache placement,'' {\em IEEE Transactions on Information
  Theory}, vol.~66, no.~3, pp.~1318--1332, 2020.

\bibitem{yu2018characterizing}
Q.~Yu, M.~A. Maddah-Ali, and A.~S. Avestimehr, ``Characterizing the rate-memory
  tradeoff in cache networks within a factor of 2,'' {\em IEEE Transactions on
  Information Theory}, vol.~65, no.~1, pp.~647--663, 2018.

\bibitem{wan2021optimal}
K.~Wan, H.~Sun, M.~Ji, D.~Tuninetti, and G.~Caire, ``On the optimal load-memory
  tradeoff of cache-aided scalar linear function retrieval,'' {\em IEEE
  Transactions on Information Theory}, vol.~67, no.~6, pp.~4001--4018, 2021.

\bibitem{ma2021general}
Y.~Ma and D.~Tuninetti, ``A general coded caching scheme for scalar linear
  function retrieval,'' {\em IEEE Journal on Selected Areas in Information
  Theory}, vol.~3, no.~2, pp.~321--336, 2022.

\bibitem{chen2016fundamental}
Z.~Chen, P.~Fan, and K.~B. Letaief, ``Fundamental limits of caching: Improved
  bounds for users with small buffers,'' {\em IET Communications}, vol.~10,
  no.~17, pp.~2315--2318, 2016.

\bibitem{gomez2018fundamental}
J.~G{\'o}mez-Vilardeb{\'o}, ``Fundamental limits of caching: Improved
  rate-memory tradeoff with coded prefetching,'' {\em IEEE Transactions on
  Communications}, vol.~66, no.~10, pp.~4488--4497, 2018.

\bibitem{tian2018caching}
C.~Tian and J.~Chen, ``Caching and delivery via interference elimination,''
  {\em IEEE Transactions on Information Theory}, vol.~64, no.~3,
  pp.~1548--1560, 2018.

\bibitem{tian2018symmetry}
C.~Tian, ``Symmetry, outer bounds, and code constructions: A computer-aided
  investigation on the fundamental limits of caching,'' {\em Entropy}, vol.~20,
  no.~8, p.~603, 2018.

\bibitem{wang2018improved}
C.-Y. Wang, S.~S. Bidokhti, and M.~Wigger, ``Improved converses and gap results
  for coded caching,'' {\em IEEE Transactions on Information Theory}, vol.~64,
  no.~11, pp.~7051--7062, 2018.

\bibitem{sengupta2017improved}
A.~Sengupta and R.~Tandon, ``Improved approximation of storage-rate tradeoff
  for caching with multiple demands,'' {\em IEEE Transactions on
  Communications}, vol.~65, no.~5, pp.~1940--1955, 2017.

\bibitem{maddah2014decentralized}
M.~A. Maddah-Ali and U.~Niesen, ``Decentralized coded caching attains
  order-optimal memory-rate tradeoff,'' {\em IEEE/ACM Transactions On
  Networking}, vol.~23, no.~4, pp.~1029--1040, 2014.

\bibitem{wei2017novel}
Y.-P. Wei and S.~Ulukus, ``Novel decentralized coded caching through coded
  prefetching,'' in {\em 2017 IEEE Information Theory Workshop (ITW)},
  pp.~1--5, IEEE, 2017.

\bibitem{reisizadeh2018erasure}
H.~Reisizadeh, M.~A. Maddah-Ali, and S.~Mohajer, ``Erasure coding for
  decentralized coded caching,'' in {\em 2018 IEEE International Symposium on
  Information Theory (ISIT)}, pp.~1715--1719, IEEE, 2018.

\bibitem{yao2024generic}
Y.~Yao and S.~A. Jafar, ``On the generic capacity of k-user symmetric linear
  computation broadcast,'' {\em IEEE Transactions on Information Theory}, 2024.

\bibitem{shanmugam2013femtocaching}
K.~Shanmugam, N.~Golrezaei, A.~G. Dimakis, A.~F. Molisch, and G.~Caire,
  ``Femtocaching: Wireless content delivery through distributed caching
  helpers,'' {\em IEEE transactions on information theory}, vol.~59, no.~12,
  pp.~8402--8413, 2013.

\bibitem{daniel2019optimization}
A.~M. Daniel and W.~Yu, ``Optimization of heterogeneous coded caching,'' {\em
  IEEE Transactions on Information Theory}, vol.~66, no.~3, pp.~1893--1919,
  2019.

\bibitem{yao2024capacity}
Y.~Yao and S.~A. Jafar, ``The capacity of 3 user linear computation
  broadcast,'' {\em IEEE Transactions on Information Theory}, vol.~70, no.~6,
  pp.~4414--4438, 2024.

\bibitem{ma2025achievablearxiv}
Y.~Ma and D.~Tuninetti, ``An achievable scheme for the k-user linear
  computation broadcast channel,'' {\em arXiv preprint arXiv:2501.12322}, 2025.

\bibitem{boerner2023access}
T.~J. Boerner, S.~Deems, T.~R. Furlani, S.~L. Knuth, and J.~Towns, ``Access:
  Advancing innovation: Nsf’s advanced cyberinfrastructure coordination
  ecosystem: Services \& support,'' in {\em Practice and experience in advanced
  research computing 2023: Computing for the common good}, pp.~173--176, 2023.

\end{thebibliography}
\end{document}